\documentstyle[preprint,aps]{revtex}

\begin{document}
\title{Violation of Bell inequalities and \\
Quantum Tomography with\\
Pure-states, Werner-states and\\
Maximally Entangled Mixed States\\
created by a Universal Quantum Entangler}
\author{M. Barbieri, F. De Martini, G. Di Nepi and P. Mataloni}
\address{Dipartimento di Fisica and \\
Istituto Nazionale per la Fisica della Materia\\
Universit\`{a} di Roma ''La Sapienza'', Roma, 00185 - Italy}
\maketitle

\begin{abstract}
Entangled pure-states, Werner-states and generalized mixed-states of any
structure, spanning a $2\times 2$ Hilbert space are created by a novel
high-brilliance universal source of polarization-entangled photon pairs. The
violation of a Bell inequality has been tested for the first time with a 
{\it pure-state}, indeed a conceptually relevant {\it ideal}\ condition, and
with Werner-states. The generalized ''maximally entangled mixed states''
(MEMS) were also synthetized for the first time and their exotic properties
investigated by means of a quantum tomographic technique. PACS\ numbers:
03.67.Mn, 03.65.Ta, 03.65.Ud, 03.65.Wj
\end{abstract}


Entanglement, ''{\it the} {\it characteristic trait of quantum mechanics'' \ 
}according to Erwin Schroedinger, is playing an increasing role in nowadays
physics \cite{1}. Indeed it is the irrevocable signature of quantum
nonlocality, i.e. the scientific paradigm today recognized as the
fundamental cornerstone of our, yet uncertain understanding of the Universe.
The striking key process, the ''{\it bolt from the blue}'' [i.e. from the
skies of Copenhagen] according to Leo Rosenfeld \cite{2}, was of course the
EPR discovery in 1935 followed by a much debated endeavour ended, in the
last few decades by the lucky emergence of\ the Bell's inequalities and by
their crucial experimental verification\ \cite{3}. In the last years the
violation of these inequalities has been tested so many times by optical
experiments that (almost) no one today challenges the reality of quantum
nonlocality. However, there still exist few crucial ''loopholes'' \ of
experimental nature that have not been adequately resolved. As far as
optical experiments is concerned, the most important one is the {\it %
quantum-efficiency }\ ''loophole'' that refers to the lack of detection of \ 
{\it all couples } of entangled photons which can be generated by the EPR
source within any elementary QED\ creation process. This effect can be due
both to the limited efficiency of the detectors \cite{4} and to the loss of
the created pairs that for geometrical reasons do not reach the detectors.
While the first contribution, expressed by the {\it detector quantum
efficiency} (DQE)\ can be attain 50\% or more, the second most important
contribution, the {\it source quantum efficiency} (SQE), has been of the
order $10^{-4}$ or less in all EPR experiments performed thus far. For
instance, in all Spontaneous Parametric Down Conversion (SPDC) experiments
the test photon pair was filtered by two narrow pinholes out of a very wide $%
\overrightarrow{k}-vector$ \ distribution spatially emitted, for any common $%
\lambda $ of the emitted particles, over one or two cones, depending on the
physical orientation of the nonlinear crystal \cite{5}. In other words, the 
{\it detected} photon pairs in {\it all }Bell inequalities tests, indeed 
{\it statistical} experiments, were in a highly {\it mixed-state, }implying
that the{\it \ }ensemble joint detection probabilities related to the {\it %
emitted} particles were not directly accessible to measurement. The lack of
detection efficiency and the {\it lack of purity} of the state of the tested
pairs were of deep concern{\it \ }to John Bell{\it \ }himself \ \cite{6} and
required the adoption of reasonable \ s{\it upplementary assumptions} \ such
as ''{\it fair sampling}'' \cite{7} and ''{\it no-enhancement}'' \cite{8}
within any acceptable interpretation of the results.

On the other hand, in modern Quantum Information (QI)\ the quest for the
conceptually appealing {\it pure}-{\it states }is{\it \ }somewhat
counterbalanced, for {\it practical} reasons, by the increasing emergence of
the {\it mixed-states }in the real world{\it . }Because of the unavoidable
effects of the decohering interactions indeed these states are today
considered the basic constituents\ of modern QI and Quantum Computation as
they limit the performance of \ all quantum communication protocols
including {\it quantum dense coding } \cite{9} and {\it quantum-state
teleportation} (QST) \cite{10}.{\it \ }It is not surprising that in the last
few years, within an endeavour aimed at the use of \ {\it mixed-states} \ as
a practical {\it resource,} an entire new branch of arduous mathematics and
topology has been created to investigate the quite unexplored theory of the 
{\it positive-maps} ({\it P-maps}) in Hilbert spaces in view of the
assessment of the ''{\it residual} {\it entanglement''} and of the
establishment of more general ''{\it state}-{\it separability}'' criteria 
\cite{11}. Very recently this ambitious study has reached results that are
conceptually relevant, as for instance the discovery of a {\it discontinuity}
in the structure of the{\it \ mixed-state }entanglement{\it . }Precisely{\it %
,} the identification of\ two classes: the {\it free}-{\it entanglement},
useful for quantum communication, and the {\it bound-entanglement}, a {\it %
non-distillable} mysterious process, elicited a fascinating new horizon
implied by the basic question: what is the role of {\it bound} entanglement
in Nature \cite{12} ?\ 

Since any real advancement in modern physics cannot be attained without
parallel, mutually testing and inspiring, experimental and theoretical
endeavours, it appears evident the present need in QI of a universal,
flexible source by which entangled {\it pure-} as well {\it mixed-states} of
any structure could be easily enginerered in a reliable and reproducible
way. In the present work we present a\ nonlinear (NL)\ optical SPDC\ source
that indeed possesses these properties. Another lucky structural property of
this source is the very {\it high-brilliance}, order of magnitude larger
than the conventional ones. Consistently with the above considerations, in
the present work this source will be applied to two different, somewhat
''extreme'' experiments, both implying a bi-partite, two qubit entangled{\it %
\ }state. A first experiment will consist of the first Bell inequality
violation experiment involving a \ {\it pure-state, }$SQE\approx 1$, i.e.%
{\it \ }by which {\it all }the SPDC\ generated photon pairs are allowed to
excite the cathode of the testing detectors. By a second experiment, the
same inequality will be tested by several Werner{\it \
transition-mixed-states,} i.e. belonging to a set for which the violation is
theoretically expected to be zero \cite{13}. In addition, a quantum
tomographic analysis will be then undertaken of \ the Werner states with
variable mixing parameters. At last, the highly relevant ''{\it maximally
entangled mixed states}'' (MEMS), today of common interest, will be created
and tested by the same technique \cite{14}. All these states have been
easily syntetized by our source for the first time, to the best of our
knowledge.\newline
Let us give here a detailed description of the apparatus. A slab of $\beta $%
-barium-borate (BBO) NL\ crystal, $.5mm$ thick and cut for Type I\
phase-matching, was excited by a vertically $(V)$ polarized, slightly
focused, cw UV $Ar^{+}$ laser beam ($\lambda _{p}=363.8nm$) with wavevector
(wv) $-{\bf k}_{p}$, i.e. directed towards the left-hand-side (lhs) of
Fig.1. In our experiment the investigated NL interaction was $\lambda $-$%
\deg $enerate, i.e. the photons of each SPDC generated $i^{th}$-pair had 
{\it equal} wavelengths (wl) $\lambda =\lambda _{ji}=727.6nm$, $(j=1,2)$,
common horizontal $(H)\ $linear-polarization $(\overrightarrow{\pi })$, and
were emitted with \ {\it equal probability, }over a corresponding pair of $\ 
$wavevectors (wv) ${\bf k}_{ji}\ $belonging to the surface of a cone with
axis ${\bf k}_{p}$ and aperture $\alpha \simeq 2.9%
{{}^\circ}%
$. The {\it product-state} of each emitted $i^{th}$-pair was expressed as: $%
\left| \Phi \right\rangle _{i}$ =$\left| HH\right\rangle _{i}$, where the
shorthand: $|XY\rangle \equiv |X\rangle \otimes |Y\rangle $ is and will be
used henceforth. The emitted radiation and the UV laser beam directed
towards the lhs of Fig.1 were then back-reflected by a multilayer-dielectric
spherical mirror $M$ \ with curvature radius $R=15cm$, highly reflecting $%
(\approx 99\%)\ $both $\lambda $ and $\lambda _{p}$, placed at a distance $%
d=R$ \ from the crystal. A zero-order $\lambda /4$ waveplate (wp) placed
between $M\ $and the NL\ crystal, i.e. in the $"d-\sec tion"$, as we call it,%
$\ $intercepted twice both back-reflected $\lambda $ and $\lambda _{p}$
beams and then rotated by $90%
{{}^\circ}%
\ $the polarization $(\overrightarrow{\pi })$ of the back-reflected photons
with wl $\lambda \ $while leaving in its original $\overrightarrow{\pi }$
state the back-reflected UV beam $\lambda _{p}\approx 2\lambda _{j}$. In
facts, it has been verified that the $\lambda /4$\ wp acted closely as a $%
\lambda _{p}/2$ wp. The\ back-reflected UV\ beam excited in the direction $%
{\bf k}_{p}$ an identical albeit distinct SPDC process with emission of a
new radiation cone directed towards the (rhs) of Fig.1 with axis ${\bf k}%
_{p} $. In this way each $i^{th}$-pair of correlated vw's, ${\bf k}_{ji}\ $%
(j=1,2)$\ $originally SPDC$\ $generated towards the lhs of Fig. 1 was made,
by optical back-reflection and a unitary $\overrightarrow{\pi }$-flipping
transformation, ''{\it in principle} {\it indistinguishable'',} i.e. for any 
{\it ideal} detector placed in{\it \ }${\cal A}$ and/or ${\cal B}$, with
another pair originally generated towards the rhs and carrying the state $%
\left| HH\right\rangle _{i}$. In virtue of \ the quantum superposition
principle, the state of \ the overall radiation, resulting from the
overlapping of \ the two {\it indistinguishable} cones expressing the
corresponding overall ${\bf k}_{ji}$-distributions, was then expressed by
the {\it pure}-{\it state}: 
\begin{equation}
|\Phi \rangle =2^{-%
{\frac12}%
}\left( |HH\rangle +e^{i\phi }|VV\rangle \right)
\end{equation}
an entangled Bell-state, with a phase $(0\leq \phi \leq \pi )$\ reliably
controlled by micrometric displacements $\Delta d\ $of $M$\ along ${\bf k}%
_{p}$. A positive lens with focal-length $F$\ transformed the overall
emission \ {\it conical }distribution\ into a {\it cylindrical }one \ with
axis ${\bf k}_{p}$. The transverse circular section of this one identified
the ''{\it Entanglement-ring''} (E-ring)\ with diameter $D=2\alpha F$. Each
couple of \ points symmetrically opposed through the center of the E-ring
were then correlated by quantum entanglement. An annular mask with diameter $%
D=1.5cm$ and width $\delta =.1cm$ provided an accurate spatial selection of
the E-ring and an efficient filtering of the unwanted UV light out of the
measurement apparatus. The E-ring spatial photon distribution was divided in
two equal portions along a vertical axis by a prism-like, two-mirror system
and collected by two lenses that focused all the radiation on the active
cathodes of two independent measurement devices at sites ${\cal A}$ and $%
{\cal B}$: {\it Alice} and {\it Bob.} Optionally, two optical fibers could
convey the radiation to two far apart stations, ${\cal A}$ and ${\cal B}$.
The indices $1$ and $2$ will also be adopted henceforth to identify local
quantities as angles, states etc. measured in spaces ${\cal A}$ and ${\cal B}
$,\ respectively. The radiation reaching sites ${\cal A}$ and ${\cal B}$ was
detected by two Si-APD photodiodes SPCM-AQR14, with $DQE=65\%$ and dark
count rate $\simeq 50s^{-1}$. Typically, two equal interference filters
(IF)\ were placed in front of the ${\cal A}$ and ${\cal B}{\bf \ }$%
detectors. \ The bandwidth $\Delta \lambda _{j}=6nm\ $of the two IF's
determined the \ {\it coherency thickness} $\delta \ \ $of\ the cylindrical
distribution and the {\it coherence-time }of the emitted photons:\ $\tau
_{coh}${\it \ }$\approx 140$ femtoseconds. \ The simultaneous detection of
the {\it whole ensemble} of the SPDC emitted entangled pairs allowed a
quantitative assesment of\ \ the absolute\ \ {\it brightness} \ of the
source. The UV pump power adopted in the experiment was: $P_{p}$ $\simeq
40mW $. At such a low power typically $4000$ coincidences per second\ were
detected, \ thus outperforming by an order at least $10^{3}\ $the overall 
{\it quantum efficiency} \ of the common SPDC\ sources \cite{5}.\ On the
other hand in these conditions the NL\ parametric gain was so small, $%
g<10^{-3}\ $that the ratio of the probabilities for simultaneous {\it %
stimulated-emission} of {\it two}$\ i^{th}$-pairs and the one for the {\it %
spontaneous-emission} of one $i^{th}$-pair was $<10^{-6}$.

The present demonstration was carried out in the $\lambda $-$\deg $enerate
condition, i.e. the highest probability process, according to NL\ Optics.
Note however that the apparatus works perfectly for a very general $\lambda $%
-{\it non-}$\deg ${\it enerate} dynamics, i.e. by allowing at the extreme
the {\it simultaneous} detection, by sufficiently broadband detectors, of
the {\it full set} of \ SPDC generated $i^{th}$-pairs with all possible
phase--matching allowed combinations of \ wl's $\lambda _{ji}$, $j=1,2$.

Note the high structural flexibility of this novel SPDC\ source. Its
structure indeed suggests the actual implementation of several relevant
schemes of quantum information and communication, including entanglement
multiplexing, joint entanglement over $\overrightarrow{\pi }$ and $k$-vector
degrees of freedom etc. Furthermore, it also suggests the realization of a
confocal cavity Optical Parametric Oscillator emitting a non-thermal, E-ring
distribution of entangled photon states. These ideas are presently being
investigated in our laboratory.

{\large Violation of Bell inequalities by a pure state: }The Bell state
expressed by Eq. 1 with $\phi =\pi $,\ i.e. a ''$\sin $glet'' over the whole 
{\it E}-{\it ring} $\ $was adopted to test the violation of a Bell
inequality by the standard coincidence technique \cite{3,15}. The adopted
angle orientations of the $\overrightarrow{\pi }$-analyzers located at the $%
{\cal A\ }$(1) and ${\cal B\ }$(2) sites were:

$\left\{ \theta _{1}=0,\theta _{1}^{\prime }=\pi /2\right\} $ and $\left\{
\theta _{2}=\pi /4,\theta _{2}^{\prime }=3\pi /4\right\} $, together with
the respective orthogonal angles: $\left\{ \theta _{1}^{\bot },\theta
_{1}^{^{\prime }\bot }\right\} $ and $\left\{ \theta _{2}^{\bot },\theta
_{2}^{^{\prime }\bot }\right\} $. By these values, the standard
Bell-inequality parameter could be evaluated \cite{3,7}:

$S$=$\left| P\left( \theta _{1},\theta _{2}\right) -P(\theta _{1},\theta
_{2}^{\prime })+P\left( \theta _{1}^{\prime },\theta _{2}\right) +P\left(
\theta _{1}^{\prime },\theta _{2}^{\prime }\right) \right| $ where:

$P\left( \theta _{1},\theta _{2}\right) $= $[C\left( \theta _{1},\theta
_{2}\right) $+$C\left( \theta _{1}^{\bot },\theta _{2}^{\bot }\right) $-$%
C\left( \theta _{1},\theta _{2}^{\bot }\right) $-$C\left( \theta _{1}^{\bot
},\theta _{2}\right) ]\times $

$[C\left( \theta _{1},\theta _{2}\right) $+$C\left( \theta _{1}^{\bot
},\theta _{2}^{\bot }\right) $+$C\left( \theta _{1},\theta _{2}^{\bot
}\right) $+$C\left( \theta _{1}^{\bot },\theta _{2}\right) ]^{-1}$and $%
C\left( \theta _{1},\theta _{2}\right) $ is the coincidence rate measured at
sites${\cal A}$ and ${\cal B}$. Fig.2 shows the$\ \overrightarrow{\pi } $%
-correlation pattern obtained\ by varying the angle $\theta _{1}$ of the $%
{\cal B\ }$analyzer in the range $(45%
{{}^\circ}%
-135%
{{}^\circ}%
)$, having kept fixed the angle of the ${\cal A}$ analyzer : $\theta _{2}=45%
{{}^\circ}%
$. The dotted line expresses in Fig.2 the limit boundary between the quantum
and the {\it ''classical''} regimes. The measured value $S=2.5564\pm .0026 $%
, obtained by integrating the data over $180s$, corresponds to a violation
as large as $213$ standard deviations respect to the limit value $S=2\ $%
implied by {\it local realistic }theories{\it \ }\cite{7}. A very small
amount of noise appears to affect in Fig.2 the experimental data which also
fit well the theoretical (continuous) curve expressing the {\it ideal}
interferometric pattern with maximum {\it visibility:} $V=1 $. In addition
to its remarkable conceptual relevance because of the adoption of a {\it %
pure-state}, this result expresses the good overall performance of our
optical system including the accuracy obtainable for the optical
superposition and focusing of large beams. The good performance was, of
course also attributable due to the {\it high brightness} of the source. In
facts, owing to $SQE\approx 1$,\ a large set of statistical data could be
accumulated in exceedingly short measurement times and with very low UV\
pump powers.

{\large Generation and characterization of \ Werner states. }Because of the
peculiar spatial superposition property of the output state, the present
apparatus appears to be an ideal source of {\it any } bi-partite, two-qubit
entangled state, either {\it pure} or {\it mixed}. In particular of \ the
Werner state: $\rho _{W}=p|\Psi _{-}\rangle \left\langle \Psi _{-}\right| +%
\frac{1-p}{4}{\bf I}$ \ consisting of a mixture of a {\it pure} singlet
state $|\Psi _{-}\rangle =2^{-%
{\frac12}%
}\left\{ \left| HV\right\rangle -\left| VH\right\rangle \right\} $ with
probability $p$ ($0\leq p\leq 1$) and of a fully {\it mixed-state }expressed
by the unit operator ${\bf I}$. The corresponding density matrix, expressed
in the basis $|HH\rangle $, $|HV\rangle $, $|VH\rangle $, $|VV\rangle $ is: 
\begin{equation}
\rho _{W}=\left( 
\begin{array}{cccc}
A & 0 & 0 & 0 \\ 
0 & B & C & 0 \\ 
0 & C & B & 0 \\ 
0 & 0 & 0 & D
\end{array}
\right)
\end{equation}

with:\ $A$=$D$=$%
{\frac14}%
(1-p)$, $B$=$%
{\frac14}%
(1+p)$, $C$=$-p/2$. The Werner states possess a highly conceptual and
historical value because, in the probability range $[1/3<p<1/\sqrt{2}]$,
they {\it do not} violate any Bell's inequality in spite of being in this
range {\it nonseparable} entangled states, precisely {\it NPT states} \cite
{11}.

How to syntezize by our source these paradigmatic, utterly remarkable states
?

Among many possible alternatives, we selected a convenient \ {\it patchwork }
technique implying the following steps: ${\bf [1]}$ Making reference to the
original {\it source-state} expressed by Eq.1, a {\it singlet }state $|\Psi
_{-}\rangle $ was easily obtained by inserting a $\overrightarrow{\pi }${\it %
-flipping,} zero-order $\lambda /2$ wp in front of detector ${\cal B}$. $[%
{\bf 2}]${\bf \ }A anti-reflection coated glass-plate ${\it G}$, $200\mu m\ $%
thick, inserted in the $d-\sec tion$ with a variable trasverse position $%
\Delta x$, introduced a decohering fixed time-delay $\Delta t>\tau _{coh}\ $%
that spoiled the{\it \ \ indistinguishability} of the {\it intercepted
portions} of the overlapping {\it quantum-interfering} radiation cones:
Fig.3, inset. As a consequence, {\it all nondiagonal} \ elements of $\rho
_{W}$ contributed by the surface sectors ${\bf B}+{\bf C}$ of the E-ring,
the ones optically intercepted by $G$, were set to {\it zero }while the non
intercepted sector ${\bf A}$ expressed the {\it pure-state} singlet
contribution to $\rho _{W}$. ${\bf [3]}$ A $\lambda /2$ wp was inserted in
the semi-cylindrical photon distribution reflected by the beam-splitting
prism towards the detector ${\cal A}$. Its position was carefully adjusted
in order to intercepts {\it half }of the ${\bf B}+{\bf C}\ $sector, i.e. by
making ${\bf B}={\bf C}$. Note that only {\it half }of the E-ring needed to
be intercepted by the optical plates, in virtue of\ \ the EPR nonlocality.
The other, optically non-intercepted, half of the E-ring is not represented
in Fig.3, inset. In summary, the sector ${\bf A} $ of the E-ring contributed
to $\rho _{W}\ \ $with a {\it pure} state $p|\Psi _{-}\rangle \left\langle
\Psi _{-}\right| $, the sector ${\bf B}+{\bf C}=2{\bf B}$ with the {\it %
statistical} {\it mixture}: $\frac{1-p}{4}\left\{ \left[ \left|
HV\right\rangle \left\langle HV\right| \text{+}\left| VH\right\rangle
\left\langle VH\right| \right] \text{\ +\ }\left[ \left| HH\right\rangle
\left\langle HH\right| \text{+}\left| VV\right\rangle \left\langle VV\right| %
\right] \right\} $ and the probability $p$, a monotonic function of $\Delta
x $,\ could be easily varied over its full range of values: $p\propto \Delta
x$ for small $p$. The two extreme cases are: ${\bf [a]}$ The ${\it G}$ and
the $\lambda /2$\ wp\ intercepting the beam towards ${\cal A\ }$are absent: $%
{\bf B}={\bf C}=0$. The $A$ section covers {\it half} of the E-ring and: $%
\rho _{W}\equiv |\Psi _{-}\rangle \left\langle \Psi _{-}\right| $; ${\bf [b]}
$ The ${\it G}$ plate intercepts {\it half} of the E-ring and the position
of the $\lambda /2$ wp intercepting the beam towards ${\cal A\ }$is set to
make ${\bf B}={\bf C}$. In this case ${\bf A}=0$ and: $\rho _{W}\equiv \frac{%
1}{4}{\bf I}$. Optionally, the setting of the $\lambda /2$ wp intercepting
the beam towards ${\cal A\ }$could be automatically activated by the {\it %
single} setting $\Delta x$, e.g. via an electromechanical servo.

{\it Any }Werner state could be realized by this technique,\ by setting $%
{\bf B}={\bf C}\ $and by adjusting the value of $\ p(\Delta x)$. Far more
generally, {\it all possible }bi-partite states in $2\times 2\ $dimensions
could be created by this technique. This indeed expresses the\ ''{\it %
universality''} of our source.

A set of Werner states was indeed syntetized and several relevant properties
investigated by our method leading to the experimental results shown in Fig
3 and 4a). \ A relevant property of any mixed-state, the ''{\it tangle}''$%
T=[C(\rho )]^{2}$, i.e. the square of \ the {\it concurrence} $C(\rho )$,$\ $%
is directly related to the {\it entanglement of formation} $E_{F}(\rho )\ $%
and expresses the degree of entanglement of $\rho $ \cite{16}. Another
important property of the mixed-states is the ''{\it linear entropy'' }$%
S_{L} $=\ $d(1-Tr\rho ^{2})/(d-1)$, $S_{L}$=$(1-p^{2})$ for Werner states,
which quantifies the degree of disorder, viz. the {\it mixedeness} of a
system with dimensions $d$ \cite{17}. In virtue of the very definition of $%
C(\rho )$, these two quantities are found to be related, for Werner states,
as follows: $T_{W}\left( S_{L}\right) =%
{\frac14}%
(1-3\sqrt{1-S_{L}})^{2}\ $for $0\leq S_{L}\leq 8/9$ \cite{16}. As shown in
Fig.3, for $S_{L}\ $in the range $[\frac{1}{2}\leq S_{L}\leq 8/9]$ \ the
Bell inequalities are not violated while $\rho _{W}\ \ $is a {\it separable}%
, PPT state for $S_{L}>8/9\ $and $T\left( S_{L}\right) =0$ \cite{11}. The
experimental result shown in Fig.4a of a standard tomographic analysis of
the Werner state corresponding to $p\simeq 0.42$ reproduces graphically, and
quite accurately the structure of the matrix $\rho _{W}$ expressed by Eq.2.
The properties $T_{W}\left( S_{L}\right) $ of a full set of Werner states
are also reported in Fig. 3. There the experimental points (full circles)
are determined according to the following procedure. First, the position $%
\Delta x\ $of the plate $G$ is set according to a pre-determined,
zeroth-order value of $p$. Then, a tomographic experiment reproducing a
result similar to Fig. 4a \ and a numerical optimization procedure lead to
the determination of the {\it actual} value of $\rho $, $Tr\rho ^{2}$ and
then of $S_{L}$. It leads to the {\it actual} value of $p$, $p_{sp}$ \cite
{18}. Finally, $\rho $ is adopted to evaluate the {\it actual} value of $%
T_{W}=[C(\rho )]^{2}$. We may check in \ Fig.3 the good agreement of the
experimental data with the theoretical result expressed by the plotted
function $T_{W}\left( S_{L}\right) $. The state represented by Fig.4a, and
also reproduced in Fig.3, {\it does not} violate any Bell inequality, in
spite of being a {\it non separable} one. Indeed, the corresponding
Bell-inequality parameter has been experimentally measured: $S=1.048\pm
0.011 $. The transition to $S>2\ $has been experimentally determined, and
found consistent with theory, by increasing the value of $p\ $in order to
set $S_{L}=(1-p^{2})<%
{\frac12}%
$.

{\large Generation and characterization of \ MEMS\ states. }As a final
demonstration of the universality of our method, a\ full set of \ ''{\it %
maximally entangled mixed states}'' (MEMS) was syntetized by our source and
tested again by quantum tomography \cite{14,18}. To the best of our
knowledge, a similar result was not previously reported in the literature.
On the other hand, according to the introductory notes expressed above, the
MEMS are to be considered, for {\it practical} reasons, as important
resources of modern QI because they achieve the {\it greatest possible }%
entanglement for a given mixedeness, i.e. the one which is the unavoidable
manifestation of {\it decoherence}. Of course, in order to syntetize the
MEMS a convenient partition of the {\it E-ring}, different from the one
shown in Fig.3, has been adopted. The class of MEMS generated and tested by
our method are expressed by $\rho _{MEMS\;}$which is again given in matrix
form by Eq. 2, with the following parameters: $A$=$(1-2g(p))$,$\ B$=$g(p)$,$%
\ C$=$-p/2$, $D$=$0\ $and: $g(p)=p/2$ for $p\geq 2/3$ and $g(p)=1/3$ for $%
p<2/3$. The tomographic result shown in Fig.4b reproduces graphically, and
with fair accuracy the $\rho _{MEMS\;}$structure with the parameter: \ $%
p=0.56$ and then: $g(p)=1/3$. However, the agreement between experimental
and theoretical results expressed in Fig.3 is less satisfactory for MEMS
than it was for Werner states. This is due to the strict experimental
requirements implied by the tricky realization of these exotic objects.
Improvements in this direction as well as investigations with the new source
in related domains of quantum information are presently\ being tackled in
our laboratory \cite{19}.

Thanks are due to W.Von Klitzing for useful discussions and early
involvement in the experiment. This work was supported by the FET European
Network on Quantum Information and Communication (Contract IST-2000-29681:
ATESIT) and by PRA-INFM\ 2002 (CLON).

\centerline{\bf Figure Captions}

\vskip 8mm

\parindent=0pt

\parskip=3mm

Figure 1: Layout of the {\it universal, high-brightness }source of \
polarization $(\overrightarrow{\pi })$ entangled {\it pure-states} \ and of
general {\it mixed-states} of \ photons.

Figure 2: Violation of \ a Bell Inequality by a ''singlet'' {\it pure-state. 
}The {\it dotted }curve expresses the transition between standard quantum
theory and \ {\it local realistic theories}. The {\it continuous} curve
expresses the {\it optimal} theoretical behavior with\ interferometric
visibility: $V=1$. The statistical noise of each experimental datum is
expressed by the corresponding error flag.

Figure 3: Experimental plots of \ the ''{\it Tangle}'' $T$\ vs the ''{\it %
Linear entropy''} $S_{L}$\ for Werner states (full circles) and {\it %
maximum-entangled-mixed-states} (MEMS) (open circles). The vertical dotted
line expresses the boundary between the two regimes corresponding to
non-separable\ states violating and non-violating Bell inequalities. \ The
shaded region on the upper right side is dynamically inaccessible. INSET:
Partition of the (half) {\it Entanglement-ring} into the spatial
contributions of the emitted entangled-pair distribution to an overall
output Werner-state.

Figure 4: Experimental quantum tomographic reconstructions of: (a) A
Werner-state $\rho _{W\;}$with ''singlet'' probability $p=0.42$ and: (b) A
MEMS-state $\rho _{MEMS\;}$with $p=0.56$. The two tomographic patterns
reproduce graphically the structure of the $\rho $-matrix expressed by Eq. 2
with the appropriate parameters.

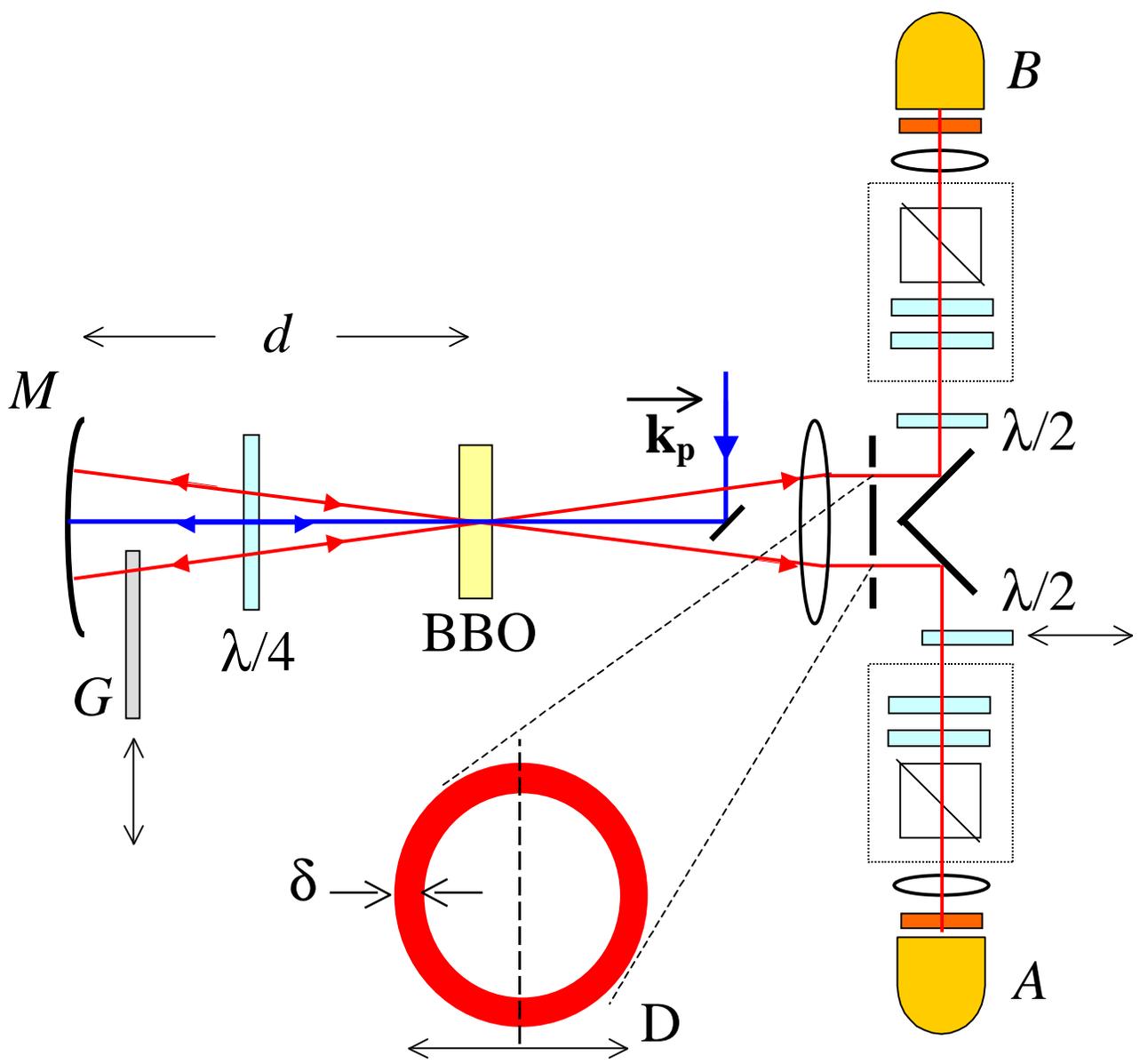
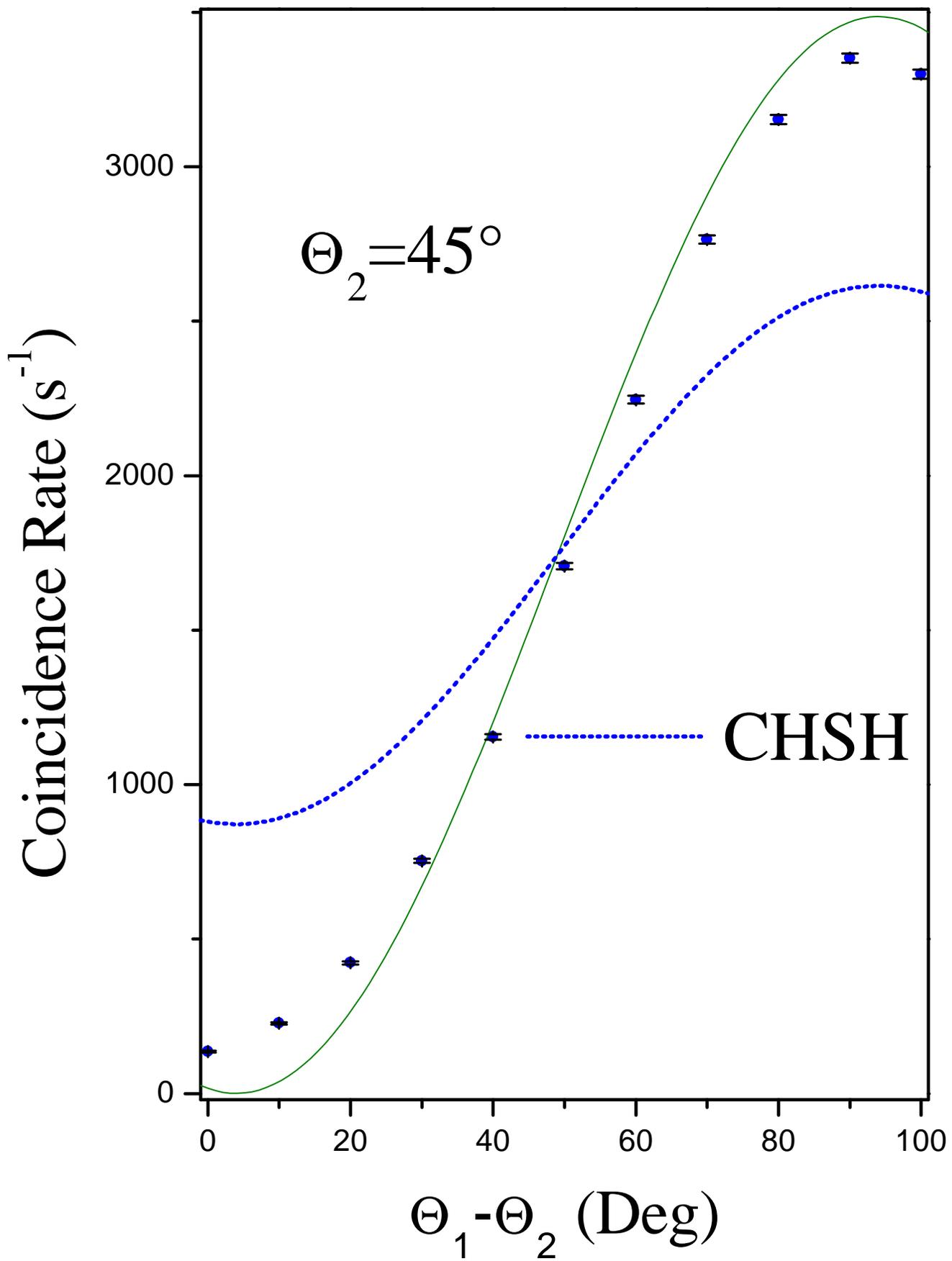
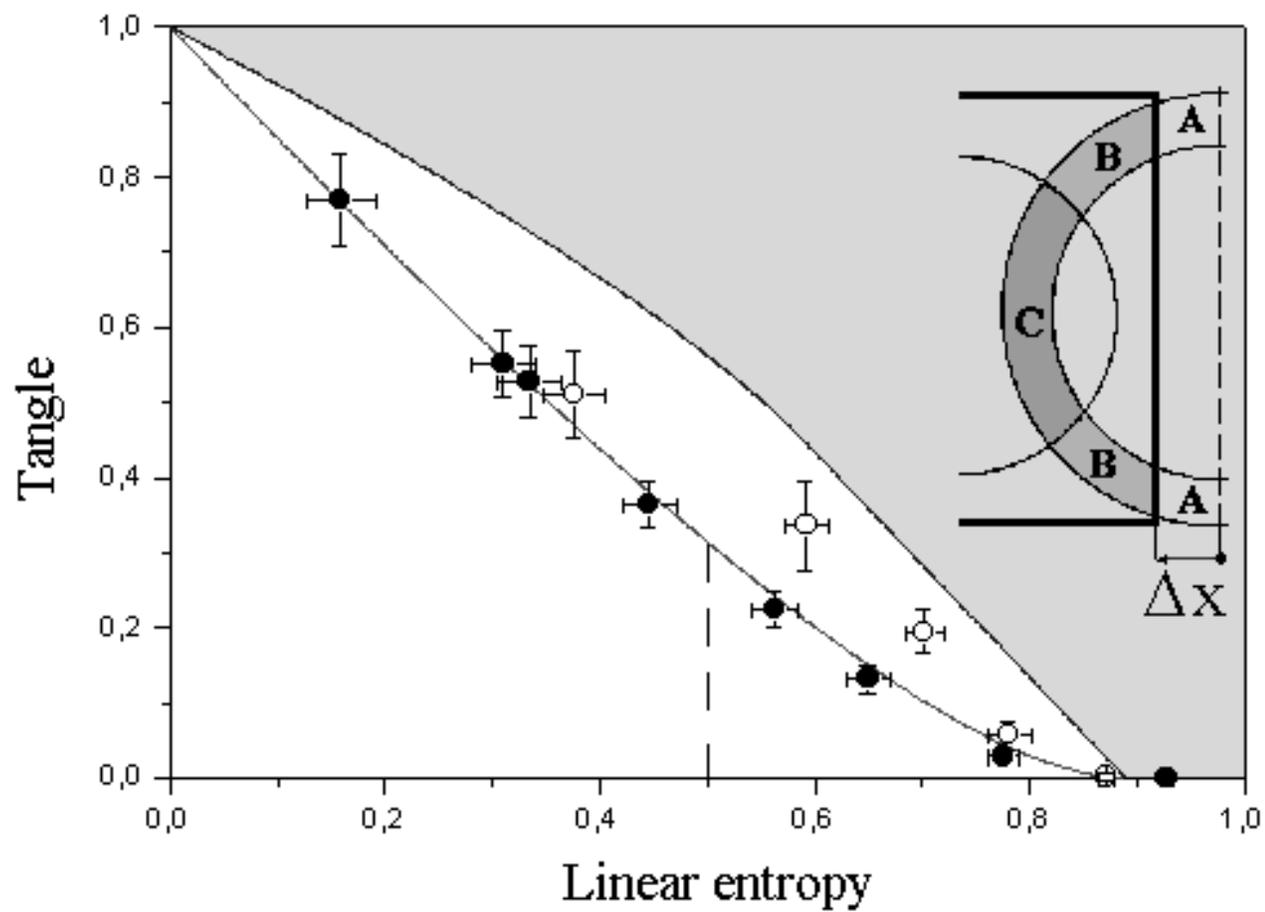
\end{document}